\documentclass[fleqn,twoside,11pt]{article}

\usepackage{amssymb}
\usepackage{graphicx}

\usepackage{amsmath}
\begin{document}
\noindent
{\sf University of Shizuoka}

\hspace*{13cm} {\large US-04-11}
\vspace{2mm}
%\draft

\begin{center}
%\title{

{\Large\bf Maximal $CP$ Violation Hypothesis\\[.1in]

and Phase Convention of the CKM Matrix}\\

\vspace{3mm}

%\author{
{\bf Yoshio Koide}

%\address{
{\it Department of Physics, University of Shizuoka, 
52-1 Yada, Shizuoka 422-8526, Japan\\
E-mail address: koide@u-shizuoka-ken.ac.jp}

%\vspace{2mm}
%\date{\today}
\end{center}

\vspace{3mm}
%\maketitle
\begin{abstract}
The maximal $CP$ violation hypothesis depends on the phase
convention of the Cabibbo-Kobayashi-Maskawa matrix.
A phase convention which leads to successful prediction 
under the maximal $CP$ violation hypothesis is searched,
and thereby, possible structures of the quark mass matrices
are speculated.
\end{abstract}

%\pacs{

%\maketitle

\vspace{5mm}
%%%%%%%%%%%%%%%%%%%%%%%%%%%%%%%%%%%%%%%%%%%%%%%%%%%%%%%%%%%%%
%\begin{multicols}{2}

%\narrowtext

\noindent{\large\bf 1 Introduction} \ 

Recent remarkable progress of the experimental $B$ physics
\cite{B} has made 
possible to 
know the magnitude of the $CP$ violation in the quark sector. 
We are interested 
what logic can give the observed magnitude of the $CP$ violation. 
For this 
subject, for example, we know an attractive hypothesis, 
the so-called 
``maximal $CP$ violation" hypothesis \cite{maxCP}. 
However, the conventional ``maximal $CP$ 
violation" hypothesis cannot give the observed magnitude of 
the $CP$ violation, 
as we discuss later.

We are also interested that, which quark mass 
matrix element, the $CP$ violation originates in 
(in other words, which of quark mass matrix elements is 
accompanied by a $CP$ violating phase). 
However, it is usually taken that this question is meaningless, 
because we know 
that the observable quantities are invariant under 
the rephasing of the Cabibbo-Kobayashi-Maskawa (CKM) \cite{Cabibbo,KM}
matrix.
For example, we cannot physically distinguish the standard CKM matrix
phase convention \cite{SDCKM}
$$
V_{SD} = R_1 (\theta_{23}) P_3(\delta_{13}) R_2 (\theta_{13}) 
P_3^{\dagger} (\delta_{13}) R_3 (\theta_{12})
$$
$$
= \left(
\begin{array}{ccc}
c_{13} c_{12} & c_{13} s_{12} & s_{13} e^{-i\delta_{13}} \\
-c_{23} s_{12} -s_{23} c_{12} s_{13} e^{i\delta_{13}} &
c_{23} c_{12} -s_{23} s_{12} s_{13} e^{i\delta_{13}} &
s_{23} c_{13} \\
s_{23} s_{12} -c_{23} c_{12} s_{13} e^{i\delta_{13}} &
-s_{23} c_{12} -c_{23} s_{12} s_{13} e^{i\delta_{13}} &
c_{23} c_{13} 
\end{array} \right) ,
\eqno(1.1)
$$
from the original CKM matrix phase convention by Kobayashi and
Maskawa (KM) \cite{KM}
$$
V_{KM} = R_1^{T} (\theta_2) P_3 (\delta_{KM} + \pi) R_3 (\theta_1)
R_1 (\theta_3)
$$
$$
= \left(	
\begin{array}{ccc}
c_{1}  & -s_{1} c_{3} & -s_{1} s_3  \\
s_1 c_2 & c_1 c_2 c_3 -s_2 s_3 e^{i\delta_{KM}} &
c_1 c_2 s_3 + s_2 c_3 e^{i \delta_{KM}} \\
s_1 s_2 & c_1 s_2 c_3 +c_2 s_3 e^{i\delta_{KM}} &
c_1 s_2 s_3 -  c_2 c_3 e^{i \delta_{KM}}
\end{array} \right) ,
\eqno(1.2)
$$
where
$$
R_1 (\theta) = \left(
\begin{array}{ccc}
1 & 0 & 0 \\
0 & c & s \\
0 & -s & c 
\end{array} \right) , \ \ \ \ 
R_2 (\theta) = \left(
\begin{array}{ccc}
c & 0 & s \\
0 & 1 & 0 \\
-s & 0 & c 
\end{array} \right) , \ \ \ \ 
R_1 (\theta) = \left(
\begin{array}{ccc}
c & s & 0 \\
-s & c & 0 \\
0 & 0 & 1 
\end{array} \right) ,
\eqno(1.3)
$$
$$
P_3 (\delta) = {\rm diag} (1, \ 1, \ e^{i \delta}), \ \ \ 
\eqno(1.4)
$$
$s = \sin \theta$ and  $c = \cos \theta$.

Although there are many different versions of the maximal $CP$ violation
hypothesis, the conventional one demands that the nature takes a value of
the $CP$ violating phase so that the rephasing invariant quantity 
\cite{J} $J$ takes its maximal value. 
In the standard CKM matrix phase convention, the quantity $J$ is given by
$$
J = c^2_{13} s_{13} c_{12} s_{12} c_{23} s_{23} \sin \delta_{13} \ \ ,
\eqno(1.5)
$$
i.e.
$$
J = \frac{|V_{11}||V_{12}||V_{33}||V_{23}||V_{13}|}{1 - |V_{13}|^2}
\sin \delta_{13} \ \ .
\eqno(1.6)
$$
The maximal $CP$ violation hypothesis demands $\sin \delta_{13} =1$, so that
we obtain
$$
J \simeq |V_{us}| |V_{cd}| |V_{ub}| \ ,
\eqno(1.7)
$$
where we have used the observed fact $1 \gg |V_{us}|^2 \gg |V_{cd}|^2 \gg
|V_{ub}|^2$. The choice $\delta_{13} = \pi /2$ also predicts
$$
|V_{td}| = \sqrt{(s_{23} s_{12})^2 + (c_{23} c_{12} s_{13})^2} 
= 0.00976 \pm 0.00016 \ , 
\eqno(1.8)
$$
$$
\alpha = {68.5^{\circ}}^{+3.2^\circ}_{-2.7^\circ}  
\simeq \sin^{-1} (|V_{us}| |V_{cb}| /|V_{td}|) \ ,
\eqno(1.9)
$$
$$
\beta = {21.5^{\circ}}^{-3.2^\circ}_{+2.7^\circ}  
\simeq \sin^{-1} (|V_{ub}| / |V_{td}|) \ ,
\eqno(1.10)
$$
$$
\gamma = 89.96^{\circ} \pm 0.00^{\circ} \simeq \sin^{-1} (1) \ ,
\eqno(1.11)
$$
where angles $\alpha , \ \beta$ and $\gamma$ are defined by
$$
\alpha = {\rm Arg} \left[-\frac{V_{31} V^{*}_{33}}{V_{11} V^{*}_{13}} \right]
= \sin^{-1} \left[\frac{|V_{12}| |V_{22}|}{|V_{31}| (1-|V_{13}|^2)} 
\sin \delta_{13} \right]
 \ ,
\eqno(1.12)
$$
$$
\beta = {\rm Arg} \left[-\frac{V_{21} V^{*}_{23}}{V_{31} V^{*}_{33}} \right]
= \sin^{-1} \left[\frac{|V_{11}| |V_{12}| |V_{13}|}
{|V_{21}| |V_{31}| (1-|V_{13}|^2)} 
\sin \delta_{13} \right]
 \ ,
\eqno(1.13)
$$
$$
\gamma = {\rm Arg} \left[-\frac{V_{11} V^{*}_{13}}{V_{21} V^{*}_{23}} \right]
= \sin^{-1} \left[\frac{|V_{12}| |V_{33}|}{|V_{21}| (1-|V_{13}|^2)} 
\sin \delta_{13} \right]
 \ ,
\eqno(1.14)
$$
and we have used the observed values \cite{PDG04}
$$
\begin{array}{c}
|V_{us}| = 0.2200 \pm 0.0026 \ , \\
|V_{cb}| = 0.0413 \pm 0.0015 \ , \\
|V_{ub}| = 0.00367 \pm 0.00047 \ .
\end{array} 
\eqno(1.15)
$$
The world average value of $\beta$ \cite{PDG04} which has been
obtained from $B_d$ decays is
$$
\sin 2\beta = 0.736 \pm 0.049 \ \ 
\left( \beta = 23.7^\circ {}^{+2.2^\circ}_{-2.0^\circ} \right) ,
\eqno(1.16)
$$
so that the prediction (1.10) is in good agreement with
the observed value. 
However, on the other hand, the best fit for the CKM parameters
\cite{PDG04} gives
$$
\gamma = 60^{\circ} \pm 14^{\circ} \ , \ \ \ 
\beta = 23.4^{\circ} \pm 2^{\circ} \ ,
\eqno(1.17)
$$
so that the prediction of $\gamma$, (1.11), is entirely in disagreement 
with the experiments.
Therefore, the maximal $CP$ violation hypothesis must be ruled out.

However, note that this maximal $CP$ violation hypothesis depends on the
phase convention of the CKM matrix. If we use the original KM phase
convention, the rephasing invariant quantity $J$ is given by
$$
J = c_1 s_1^2 c_2 s_2 c_3 s_3 \sin \delta_{KM} \ ,
\eqno(1.18)
$$
i.e.
$$
J = \frac{|V_{11}||V_{12}||V_{13}||V_{21}||V_{31}|}{1 - |V_{11}|^2}
\sin \delta_{SD} \ ,
\eqno(1.19)
$$
and the requirement $\delta_{KM}=\pi/2$ predicts
$$
J \simeq |V_{ub}| |V_{td}| \ ,
\eqno(1.20)
$$
$$
|V_{ub}| = s_1 s_2 \simeq |V_{us}| |V_{cb}| \sqrt{1- \xi^2} \ ,
\eqno(1.21)
$$
where
$$
\xi = |V_{ub}| / |V_{us}| |V_{cb}| \ .
\eqno(1.22)
$$
(The relations between $V_{SD}$ and $V_{KM}$ can, for instance, 
be found in Ref.\cite{KN95}.)
From the observed values (1.15), we obtain the numerical results
$$
|V_{td}| = 0.0084 \pm 0.0005  ,
\eqno(1.23)
$$
$$
\alpha = 89.96^{\circ} \pm 0.00^{\circ}  ,
\eqno(1.24)
$$
$$
\beta = 23.2^{\circ}{}^{-3.8^\circ}_{+3.5^\circ}  ,
\eqno(1.25)
$$
$$
\gamma = 66.8^{\circ}{}^{+3.8^\circ}_{-3.5^\circ} .
\eqno(1.26)
$$
These results are in good agreement with the observed values (1.16)
and (1.17).

Thus, the results from the maximal $CP$ violation hypothesis depend on
the phase convention. 
(Note that we have applied the maximal $CP$ violation hypothesis to
the CKM phase convention $V_{KM}$, (1.2),
under the rotation parameters fixed.
If we apply the hypothesis to $V_{KM}$ under 
$|V_{us}|$, $|V_{cb}|$ and $|V_{ub}|$ fixed, the results are
same as in the standard phase convention.)
Such phase-convention dependence, in spite of the 
rephasing invariance of the CKM matrix, is due to that we tacitly assume
that only the phase parameter $\delta_{13}$ ($\delta_{KM}$) is free and 
it is independent of the rotation parameters $s_{ij}$ ($s_i$).

In the present paper, we systematically investigate whether there is other
phase convention which gives successful predictions or not, and we will find
an interesting phase convention which speculates successful relations for quark
masses $m_{qi}$ and the CKM matrix elements $|V_{ij}|$.
%%%%%%%%%%%%%%%%%%%%%%%%%%%%%%%%%%%%%%%%%%%%%
\vspace{2mm}

\noindent{\large\bf 2\ \  Phase conventions and the expressions of $J$} \ 

Let us give the CKM matrix $V$ as
$$
V = V(i,k) \equiv R^T_i P_j R_j R_k \ \ \ \ \ (i \neq j \neq k) ,
\eqno(2.1)
$$
where $R_i$  $(i=1,2,3)$ are defined by Eqs.~(1.3), and $P_i$ are given
by $P_1 = {\rm diag} (e^{i \delta}, \ 1, \ 1)$, \ 
$P_2 = {\rm diag} (1, \ e^{i \delta}, \ 1)$, and 
$P_3 = {\rm diag} (1, \ 1, \ e^{i \delta})$, we can show that the magnitudes of
the CKM matrix elements, $|V_{i1}|$, \ $|V_{i2}|$, \ $|V_{i3}|$, \ $|V_{1k}|$, 
\ $|V_{2k}|$ and $|V_{3k}|$, do not depend on the phase parameter $\delta$,
and the rephasing invariant quantity $J$ is given by
$$
J=\frac{|V_{i1}||V_{i2}||V_{i3}||V_{1k}||V_{2k}||V_{3k}|}
{(1 -|V_{ik}|^2 ) |V_{ik}|} \sin \delta \ .
\eqno(2.2)
$$
Note that the expression (2.2) is only dependent on $i$ and $k$, and it is 
independent of $j$ . 
Therefore, we have nine cases of $V(i,k)$. 
(This has been pointed out by Fritzsch and Xing \cite{FX-9V}.)
The expressions $V(1,3)$ and $V(1,1)$
correspond to the standard and original KM phase conventions, respectively.

For the observed fact $1 \gg |V_{us}|^2 \simeq |V_{cd}|^2 \gg |V_{cb}|^2
\simeq |V_{ts}|^2 \gg |V_{ub}|^2$, the results (2.2) are approximately given
as follows:
$$
J \simeq |V_{us}||V_{cb}||V_{ub}| \sin \delta \ ,
\eqno(2.3)
$$
for $V(1,2)$, \ $V(1,3)$, \ $V(2,1)$ and $V(2,3)$
$$
J \simeq |V_{ub}||V_{td}| \sin \delta \ ,
\eqno(2.4)
$$
for $V(1,1)$ and $V(3,3)$
$$
J \simeq |V_{cb}|^2 \sin \delta \ ,
\eqno(2.5)
$$
for $V(2,2)$, and 
$$
J \simeq |V_{us}||V_{cb}||V_{td}| \sin \delta \ ,
\eqno(2.6)
$$
for $V(3,1)$ and $V(3,2)$. The cases which can give reasonable predictions for
unitary triangle under the maximal $CP$ violation hypothesis are 
only the cases $V(1,1)$ and $V(3,3)$.

The explicit expression of $V(1,1)$ has already been given by Eq.~(1.2).
The explicit expression of $V(3,3)$ is given by
$$
V(3,3) = R_3^T (\theta_{12}^u) P_1 (\delta) R_1 (\theta_{23}) 
R_3(\theta^d_{12})
$$
$$
= \left(
\begin{array}{ccc}
e^{i \delta} c^u_{12} c^d_{12} + c_{23}s^u_{12} s^d_{12} 
&e^{i \delta} c^u_{12} s^d_{12}- c_{23} s^u_{12} c^d_{12} 
& -s_{23} s^u_{12}  \\
e^{i \delta} s^u_{12} c^d_{12} - c_{23} c^u_{12} s^d_{12} &
e^{i \delta} s^u_{12} s^d_{12} + c_{23} c^u_{12} c^d_{12} & 
s_{23} c^u_{12} \\
-s_{23} s^d_{12} & -s_{23} c^d_{12} & c_{23}
\end{array} \right) ,
\eqno(2.7)
$$
which has been proposed by Fritzsch and Xing \cite{V33}.
For the expression (2.7), we obtain the expression of $J$
$$
J = c_{23} s^2_{23} c^u_{12} s^u_{12} c^d_{12} s^d_{12} \sin \delta 
=
\frac{|V_{13}||V_{23}||V_{33}||V_{32}||V_{31}|}{1-|V_{33}|^2} \sin \delta \ ,
\eqno(2.8)
$$
and the relations
$$
\frac{s^u_{12}}{c^u_{12}} = \frac{|V_{ub}|}{|V_{cb}|} \ ,
\eqno(2.9)
$$
$$
\frac{s^d_{12}}{c^d_{12}} = \frac{|V_{td}|}{|V_{ts}|} \ ,
\eqno(2.10)
$$
$$
s_{23} = \sqrt{|V_{cb}|^2 + |V_{ub}|^2} \ .
\eqno(2.11)
$$
Under the maximal $CP$ violation hypothesis, since the matrix element
$|V_{us}|$ is given
$$
|V_{us}| = \sqrt{(c^u_{12} s^d_{12})^2 + (c_3 s^u_{12} c^d_{12})^2} \ ,
\eqno(2.12)
$$
the value of $s^d_{12}$ can be fixed by the observed values of $|V_{us}|$, \ 
$|V_{cb}|$ and $|V_{ub}|$. It is approximately given by 
$$
s^d_{12} \simeq |V_{us}| \sqrt{1 -\xi^2} \ ,
\eqno(2.13)
$$
where $\xi$ is defined by Eq.~(1.22). When we use the observed values of
$|V_{us}|$, $|V_{cb}|$ and $|V_{ub}|$, (1.15), the numerical predictions
without approximation are as follows:
$$
J = (3.01^{-0.22}_{+0.10}) \times 10^{-5}  , 
\eqno(2.14)
$$
$$
|V_{td}| = 0.00842 \pm 0.00052  ,
\eqno(2.15)
$$
$$
\alpha = 88.95^\circ {}^{+0.14^\circ}_{-0.12^\circ}  ,
\eqno(2.16)
$$
$$
\beta = 23.2^\circ {}^{-3.8^\circ}_{+3.5^\circ}  ,
\eqno(2.17)
$$
$$
\gamma = 67.8^\circ {}^{+2.7^\circ}_{-4.4^\circ}  .
\eqno(2.18)
$$
These numerical results are approximately same as those in the
original KM phase convention, but are slightly different from
the results (1.8)--(1.11).

%%%%%%%%%%%%%%%%%%%%%%%%%%%%%%%%%%%%%%%%%%%%%
\vspace{2mm}

\noindent{\large\bf 3\ \  Speculation on the quark mass matrix form} \ 

In the maximal $CP$ violation hypothesis, we have, so far, assumed that
the rotation parameters are fixed and only free parameter is the $CP$
violation phase $\delta$. This suggests the following situation.
The phase factors in the quark mass matrices $M_f$ \ $(f=u,d)$ are 
factorized by the phase matrices $P_f$ as
$$
M_f = P^{\dagger}_{fL} \widetilde{M}_f P_{fR} \ ,
\eqno(3.1)
$$
where $P_f$ are phase matrices and $\widetilde{M}_f$ are real matrices. The
real matrices $\widetilde{M}_f$ are diagonalized by rotation (orthogonal)
matrices $R_f$ as
$$
R^{\dagger}_f \widetilde{M}_f R_f = D_f 
\equiv {\rm diag} (m_{f1}, \ m_{f2}, \ m_{f3} ),
\eqno(3.2)
$$
(for simplicity, we have assumed that $M_f$ are Hermitian or symmetric matrix,
i.e. $P_{fR} = P_{fL}$ or $P_{fR} = P_{fL}^{\dagger}$ respectively), so that 
the CKM matrix $V$ is given by
$$
V = R^T_u P R_d \ ,
\eqno(3.3)
$$
where $P = P^{\dagger}_{uL} P_{dL}$. 
The quark masses $m_{fi}$ are only 
determined by $\widetilde{M}_f$. 
In other words, the rotation parameters are
given only in terms of the quark mass ratios, and independent 
of the $CP$ violating phases.
In such a scenario, the maximal $CP$ violation hypothesis means 
that the $CP$ violation parameter
$\delta$ takes its maximum value without changing the quark mass values.

For example, the choices of the standard and original KM phase conventions
suggest the quark mass matrix structures
$$
\begin{array}{c}
\widetilde{M}_u = R_2(\theta^u_{13}) R_1(\theta_{23}) D_u R^T_1(\theta_{23})
R_2^T (\theta^u_{13}) \ , \\
\widetilde{M}_d = R_2(\theta^d_{13}) R_3(\theta_{12}) D_d R^T_3(\theta_{12})
R_2^T (\theta^d_{13}) \ ,
\end{array}
\eqno(3.4)
$$
with $\theta_{13}=\theta^d_{13}-\theta^u_{13}$ and
$$
\begin{array}{c}
\widetilde{M}_u = R_3(\theta^u_{1}) R_1(\theta_{2}) D_u R^T_1(\theta_{2})
R_2^T (\theta^u_{1}) \ , \\
\widetilde{M}_d = R_3(\theta^d_{1}) R_1(\theta_{3}) D_d R^T_1(\theta_{3})
R_3^T (\theta^d_{1}) \ ,
\end{array}
\eqno(3.5)
$$
with $\theta_{1}=\theta^d_{1}-\theta^u_{1}$, respectively. 
The success of the maximal $CP$ violation hypothesis, 
(1.23)--(1.26), suggest that the mass matrix structure (3.5) is preferable
to the structure (3.4). 
However, another candidate of $V$ which 
gives the magnitude of $J$, (2.4), also gives successful results 
(2.14)--(2.18).  The case $V(3,3)$ suggests the following quark mass matrix
structure:
$$
\begin{array}{c}
\widetilde{M}_u = R_1(\theta^u_{23}) R_3(\theta^u_{12}) D_u 
R^T_3(\theta^u_{12}) R_1^T (\theta^u_{23}) \ , \\
\widetilde{M}_d = R_1(\theta^d_{23}) R_3(\theta^d_{12}) D_d R^T_3
(\theta_{12}^d) R_1^T (\theta^d_{23}) \ ,
\end{array}
\eqno(3.6)
$$
with $\delta = \delta_d - \delta_u$ and
$\theta_{23} = \theta^d_{23} - \theta^u_{23}$. The mass matrix structure
(3.6) is explicitly given by the form
$$
\widetilde{M}_f = \left(
\begin{array}{ccc}
m_{f1} c^{f2}_{12} + m_{f2} s_{12}^{f2} &
(m_{f2} - m_{f1}) c^f_{12} s^f_{12} c^f_{23} &
-(m_{f2} - m_{f1}) c^f_{12} s^f_{12} s^f_{23}  \\
(m_{f2} - m_{f1}) c^f_{12} s^f_{12} c^f_{23} &
(m_{f1} s^{f2}_{12} + m_{f2} c^{f2}_{12})c_{23}^{f2} + m_{f3} s^{f2}_{23} &
(m_{f3} - m_{f2} c_{12}^{f2} - m_{f1} s_{12}^{f2})c^f_{23} s^f_{23} \\
-(m_{f2} - m_{f1}) c^f_{12} s^f_{12} s^f_{23} &
(m_{f3} - m_{f2} c_{12}^{f2} - m_{f1} s_{12}^{f2})c^f_{23} s^f_{23} &
(m_{f1} s^{f2}_{12} + m_{f2} c^{f2}_{12})s_{23}^{f2} + m_{f3} c^{f2}_{23}
\end{array} \right) .
\eqno(3.7)
$$
In the mass matrix (3.7), the ansatz $\widetilde{M}^d_{11} =0$ leads to 
the well--known relation \cite{Vus}
$$
|V_{us}| \simeq s^d_{12} \simeq \sqrt{\frac{m_d}{m_s}} \simeq 0.22 \ .
\eqno(3.8)
$$
On the other hand, in the mass matrix structure (3.5), 
there is no simple relation such as
(3.8). Therefore, the mass matrix structure (3.6) 
[i.e. (3.7)] [and also the phase convention (2.7) ] is more attractive
to us compared with the alternative one (1.2) (the original KM phase
convention).
Furthermore, in the mass matrix (3.7),
if we assume $\widetilde{M}^u_{11} = 0$ analogous to $\widetilde{M}^d_{11} =0$,
we obtain
$$
\frac{s^u_{12}}{c^u_{12}} \simeq  
\sqrt{\frac{m_u}{m_c}} = 0.059 \ ,
\eqno(3.9)
$$
where quark mass values \cite{q-mass} at $\mu = m_Z$ have been used.
Compared with the experimental value of $|V_{ub}|/|V_{cb}|$
$$
\frac{s^u_{12}}{c^u_{12}} = \frac{|V_{ub}|}{|V_{cb}|} = 
0.089^{+0.015}_{-0.014} \ ,
\eqno(3.10)
$$
the prediction (3.9) is slightly small. 
However, this discrepancy should not be taken seriously,
because the present speculation on the quark mass matrices is
made only for main framework of the mass matrices.
The purpose of the present paper is to investigate a possible
phase convention form which can give successful predictions
for the shape of the unitary triangle under the maximal $CP$
violation hypothesis, and not to find a phenomenologically
successful quark mass matrix form, we do not go into the
phenomenology of the mass matrix form (3.7) any more.

%%%%%%%%%%%%%%%%%%%%%%%%%%%%%%%%%%%%%%%%%%%%%
\vspace{2mm}

\noindent{\large\bf 4 \ \ Conclusion} \ 

The predictions from the maximal $CP$ violation hypothesis depend on the phase
conventions of the CKM matrix $V$. 
We have systematically investigated whether
the hypothesis can give successful predictions for the magnitude of the 
rephasing invariant quantity $J$ and the shape of the unitary triangle or not.
In conclusion, we have found that, of the nine possible phase conventions 
$V(i, \ k) = R^T_i P_j R_j R_k$, only two, $V(1,1)$ (the original KM phase
convention) and $V(3,3)$ (the Fritzsh--Xing phase convention), can yield 
successful predictions.

Furthermore, we have speculated possible quark matrix forms which are suggested
from the expressions $V(i, \ k)$.
Since a texture-zero requirement $M^d_{11} =0$ in the mass matrix form
(3.7) can lead to the 
well--known relation $|V_{us}| \simeq {m_d / m_s}$, the new phase
convention $V(3,3)$ is very attractive to us rather than the original KM phase
convention $V(1,1)$.
(Of course, for experimental data analysis, 
the standard phase convention $V(1,3)$ [i.e. (1.1)] is the most 
useful expression.  
Only for discussing the relations between
the CKM matrix and the quark mass matrix forms $M_f$, the expression 
$V(3,3)$ [i.e. (2.7)] will be useful.)

Of course, we cannot ruled out a possibility that 
the maximal $CP$ violation hypothesis is not true.  
Then, from the view point of a simple texture-zero ansatz, 
the phase convention $V(2,3)$ is also attractive to us,
because the case suggests the quark mass matrix structure
$\widetilde{M}_u = R_1^u R_2^u D_u R_2^{uT} R_1^{uT}$ and
$\widetilde{M}_d = R_1^d R_3^d D_d R_3^{uT} R_1^{uT}$.
The texture-zero requirements $\widetilde{M}_{11}^u =0$ and
$\widetilde{M}_{11}^d =0$ predicts
$|V_{ub}|\simeq \sqrt{m_u/m_t}=0.0036$ and
$|V_{us}|\simeq \sqrt{m_d/m_s}=0.22$, respectively.
Those predictions are in good agreement with the observed values
(1.15).

If we apply the mass matrix structure (3.7) to the lepton sector, 
we obtain
$$
|U_{e3}| \simeq \frac{1}{\sqrt{2}} \sqrt{\frac{m_e}{m_\mu}} =0.049,
\eqno(4.1)
$$
for the $V(3,3)$ model, while
$$
|U_{e3}| \simeq \frac{1}{\sqrt{2}} \sqrt{\frac{m_e}{m_\tau}} =0.012,
\eqno(4.2)
$$
for the $V(2,3)$ model, where we have taken $s_{23}=c_{23}=1/\sqrt{2}$
from the observed fact \cite{atm,K2K} $\sin^2 2\theta_{atm} \simeq 1$.
If a near future experiment confirms the relation (4.1), the $V(3,3)$
model which is suggested from the maximal $CP$ violation hypothesis
will become promising.

%%%%%%%%%%%%%%%%%%%%%%%%%%%%%%%%%%%%%%%%%%%%%%%%
\vspace{7mm}

\centerline{\large\bf Acknowledgment}

This work was supported by the Grant-in-Aid for
Scientific Research, the Ministry of Education,
Science and Culture, Japan (Grant Number 15540283).

%%%%%%%%%%%%%%%%%%%%%%%%%%%%%%%%%%%%%%%%%%%%

\end{document}